\documentclass[conference]{IEEEtran}
\usepackage[utf8]{inputenc}

\usepackage[dvipsnames]{xcolor}
\usepackage{graphicx}
\usepackage{cite}
\usepackage{amssymb,amsmath,amsthm,amsfonts}
\usepackage{subcaption}

\usepackage[linesnumbered,ruled,vlined,noend]{algorithm2e}

\makeatletter
\newcommand{\nosemic}{\renewcommand{\@endalgocfline}{\relax}}% Drop semi-colon ;
\newcommand{\dosemic}{\renewcommand{\@endalgocfline}{\algocf@endline}}% Reinstate semi-colon ;
\let\oldnl\nl% Store \nl in \oldnl
\newcommand{\nonl}{\renewcommand{\nl}{\let\nl\oldnl}}% Remove line number for one line
\makeatother
\usepackage{booktabs}

\newcommand{\ie}{\textit{i.e.},}

\newcommand{\cat}[1]{\medskip\noindent\textbf{#1.}}

\newcommand{\ourAlgName}{DeViNE}
\newcommand{\ourAlg}{\ensuremath{\mathrm{\ourAlgName}}}
\newcommand{\ourAlgFull}{DeViNE: A Decentralized Virtual Network Embedding Algorithm}

\usepackage[inline]{enumitem}

\newcommand{\mc}[1]{\mathcal{#1}}

\title{\ourAlgFull}

\author{
\IEEEauthorblockN{Farzad Habibi}
\IEEEauthorblockA{\textit{Department of Computer Science} \\
    \textit{University of California, Irvine}\\
    \textit{habibif@uci.edu}}
\and
\IEEEauthorblockN{Juncheng Fang}
\IEEEauthorblockA{\textit{Department of Computer Science} \\
    \textit{University of California, Irvine}\\
    \textit{junchf1@uci.edu}}

}

\IEEEoverridecommandlockouts
\begin{document}

\maketitle

% \begin{abstract}
% Virtual Network Embedding (VNE) is an essential component of network virtualization technology. 
% %
% Prior works on VNE mainly focused on resource efficiency and did not address the scalability as a first-grade objective. Consequently, the ever-increasing demand and size render them less-practical.
% %
% The few existing designs for mitigating this problem either do not extend to multi-resource settings or do not consider the physical servers and network simultaneously.
% %
% In this work, we develop \ourAlg, a parallelizable VNE solution based on spatial Graph Neural Networks (GNN) that clusters the servers to guide the embedding process towards an improved runtime and performance. 
% %
% Our experiments using simulations show that the parallelism of \ourAlg\ reduces its runtime by a factor of $8$. Also, \ourAlg\ improves the revenue-to-cost ratio by about $18\%$, compared to other simulated algorithms.
% \end{abstract}

\begin{abstract}
Virtual Network Embedding (VNE) is a technique for mapping virtual networks onto a physical network infrastructure, enabling multiple virtual networks to coexist on a shared physical network. 
Previous works focused on implementing centralized VNE algorithms, which suffer from lack of scalability and robustness. 
This project aims to implement a decentralized virtual network embedding algorithm that addresses the challenges of network virtualization, such as scalability,  single point of failure, and DoS attacks.
The proposed approach involves selecting L leaders from the physical nodes and embedding a virtual network request (VNR) in the local network of each leader using a simple algorithm like BFS. 
The algorithm then uses a leader-election mechanism for determining the node with the lowest cost and highest revenue and propagates the embedding to other leaders. 
By utilizing decentralization, we improve the scalability and robustness of the solution. 
Additionally, we evaluate the effectiveness of our fully decentralized algorithm by comparing it with existing approaches. Our algorithm performs $12\%$ better in terms of acceptance rate and improves the revenue-to-cost ratio by roughly $21\%$ to compared approaches. 
\end{abstract}
\section{Introduction}
Virtual Network Embedding (VNE) is a crucial aspect of network virtualization technology. It involves the on-demand embedding of Virtual Network Requests (VNRs) within Physical Networks (PNs) by mapping virtual nodes and links with specific resource requirements onto physical servers and paths with finite capacities.
This process is essential in providing the necessary flexibility and isolation required for deploying various network applications in a shared infrastructure.

The embedding strategy affects resource utilization, subsequently impacting the revenue and cost of the operational network. Given the NP-hard characteristic of the problem~\cite{vne_nphard}, the development of efficient embedding algorithms has been a central focus of comprehensive research~\cite{survey_1}.

However, a significant portion of existing studies face limitations regarding scalability and robustness. Most of them are centralized solutions, which encounter difficulties in scaling as the number of Virtual Network Requests (VNRs) increases. These solutions are also vulnerable to the failure of nodes responsible for the embedding process, restricting their practical use in modern, time-sensitive environments with growth demands.

A potential solution to these challenges involves the use of a decentralized network embedding algorithm, which could improve the scalability and robustness of existing solutions.

In this work, we employ a distributed algorithm, ring leader election in particular, to formulate a decentralized Virtual Network Embedding (VNE) algorithm.  This algorithm doesn't depend on a central node, instead utilizing multiple physical servers to embed virtual networks (VN) efficiently. Users can submit their VN to any physical server in the network for embedding. Each server will then run a basic breadth-first search (BFS)-like algorithm for local embedding. Subsequently, the server sends the Virtual Network Request (VNR) to other geo-distributed servers to conduct their local embedding algorithm. Ultimately, the embedding with the lowest cost and highest revenue is selected as the final embedding through a straightforward leader election algorithm.

To the best of our knowledge, \ourAlg~ is the first fully decentralized Virtual Network Embedding (VNE) algorithm documented in the literature.
\section{Motivation}
\label{sec:motivation}
Virtual Network Embedding (VNE) has widespread applications in various fields, including cloud computing, data centers, and software-defined networks~\cite{survey_1}.
Traditionally, VNE is a centralized algorithm where a single controller handles all virtual network requests. However, as the physical network size has increased, centralized VNE has struggled to process requests in real time. This scalability issue has led to the proposition of distributed solutions.

These distributed solutions segment the network into smaller partitions, applying the embedding algorithm to each to minimize the challenge of searching the entire network. Despite their advantages, these solutions still require a centralized controller to assign requests to the sub-coordinators of each partition. While this central control requirement is suitable for some environments, such as data centers, it doesn't suit others, like edge IoT networks and mobile networks.

There are three key challenges in such environments that the current distributed VNE algorithms fail to address:

\begin{enumerate}
\item \textbf{Scalability:} Besides the increasing size of the physical networks, the number of requests is also growing, and the VNR can be sent from a wide range of physical locations across the world.
\item \textbf{Robustness:} A centralized controller has inherent vulnerabilities, including susceptibility to a single point of failure and Denial of Service (DoS) attacks.
\item \textbf{Privacy:} Users typically don't want their VNRs to be accessible to the entire network or a centralized authority.
\end{enumerate}

The objective of this paper is to develop a decentralized VNE algorithm designed to effectively overcome these challenges.
\section{Related Work}
We review related works in the literature by categorizing them into two groups. See~\cite{survey_1} for a recent survey.

\subsection{Centralized VNE}

\cat{Generic VNE Algorithms}
In~\cite{chowdhury2011vineyard}, a heuristic approach based on linear programming and rounding was introduced. While their results regarding acceptance and the revenue-to-cost ratio were satisfactory, the algorithm suffered from a relatively high runtime.
To tackle the problem's scale and complexity, authors in~\cite{Kibalya_2020_CN} used dynamic programming principles. They decomposed virtual networks into a collection of edge-disjoint path segments and leveraged a multi-layer graph transformation to embed the resulting segments.
In~\cite{Cao_2020_WCNCW}, the authors considered a multi-dimensional scenario where each physical node and link is associated with a security feature. They implemented a greedy embedding approach to allocate resources while minimizing the likelihood of malicious attacks targeting virtual networks.
The authors in~\cite{Pentelas_2020_NOMS} discussed a multi-dimensional scenario in which each physical node and the link is associated with a security feature. They utilized a greedy embedding approach to allocate resources, aiming to minimize the potential of malicious attacks on virtual networks.
%
% In~\cite{Hosseini_2019_TNSM}, the authors focused on ensuring end-to-end delays for virtual links by representing the latency in physical links as random variables with known mean and variance.

\cat{Learning-based VNE Algorithms}
% In~\cite{graphNN_RL}, the authors utilized an asynchronous advantage actor-critic algorithm to automate the embedding process using exploration-exploitation techniques, where a spectral-based convolutional graph neural network is employed to extract features from the physical network that model the environment.
% 
To address the temporal dependency of the physical network state, which changes after serving requests, the authors in~\cite{Yao_2020_TNSM} framed the node embedding task as a time-series problem. They trained a recurrent neural network using the seq2seq model to learn the embedding location for virtual nodes.
In another study, referred to as GraphViNE~\cite{habibi2020graphvine}, the authors proposed a Graph Neural Network (GNN)-based approach that incorporates a BFS-like algorithm for solving the Virtual Network Embedding (VNE) problem. This method benefits from the ability of GNNs to capture complex relationships within both virtual and substrate networks. Despite using GPU resources to manage the runtime of the VNE algorithm, the runtime remains higher than many other methods.
In some works~\cite{graphNN_RL, dolati2019deepvine, elkael2022monkey}, reinforcement learning (RL) has been employed to automate the embedding process using exploration-exploitation techniques. In a particular study~\cite{elkael2022monkey}, the authors combined the Rollout Policy Adaptive Algorithm (NRPA) with a neighborhood search, proposing a Neighborhood Enhancement Policy Adaptive (NEPA) algorithm. This technique explores the policy branch of the search tree to find the optimal embedding.

All of the aforementioned studies adopt centralized approaches, with a single node responsible for embedding. However, centralized methods encounter two potential issues. First, their scalability is inherently limited, given the potential for the physical network to comprise over a million nodes. Second, the inability of multiple nodes to handle Virtual Network (VN) requests means that service providers must process these requests sequentially, not concurrently.

\subsection{Distributed VNE}
The first fully distributed VNE approach, named ADVNE~\cite{houidi2008distributed}, breaks each virtual network into several hub-and-spoke clusters, each of which is assigned to a substrate node in the physical network for the embedding task. Through the use of multi-agent systems, these clusters are embedded into substrate nodes in a distributed manner. However, due to the autonomous nature of each substrate node in ADVNE, unavoidable message overhead exists among the substrate nodes. As the size of the substrate network grows, the message overhead increases exponentially, potentially reducing ADVNE's efficiency and scalability in large-scale settings. Moreover, ADVNE doesn't consider embedding quality (like embedding costs) and substrate network constraints (like CPU and bandwidth limitations), thereby limiting its practicality in real-world applications.

To reduce message overhead and improve embedding quality, the authors in~\cite{beck2015distributed} introduced DPVNE. In this approach, the substrate network is first hierarchically divided into sub-SNs, which are organized as a binary tree. Specific substrate nodes are then selected as delegation nodes, responsible for managing external VN requests from customers. These delegation nodes receive VN requests and assign them to sub-SNs based on heuristic information. DPVNE employs a heuristic algorithm to embed VNs within the assigned sub-SNs, with the embedding process executed by the embedder node in each sub-SN. This method allows for the concurrent processing of multiple VN requests since several delegation nodes are selected. However, it doesn't suggest a method for communication between delegation nodes to select the best embedding.
Observing the promising performance of metaheuristic approaches in VNE, the authors in~\cite{song2019distributed} integrated metaheuristics into distributed VNE problems. They merged a distributed VNE system with their proposed distributed VNE algorithm to enhance the performance of distributed approaches.

While distributed algorithms have enhanced the performance and scalability of Virtual Network Embedding (VNE), they still encounter a lack of robustness due to the reliance on a standalone controller for VNE assignment. This approach introduces a single point of failure, compromising the system's resilience and scalability.

\section{Design}
In this section, we will discuss the system model of {\ourAlgName} and the detail of the decentralized virtual network embedding algorithm.

% The physical network is represented as an un-directed graph $G_{p}=(V_{p}, E_{p})$. The set of $N$ physical nodes $V_{p}=\{v^{1}_{p},\dots,v^{N}_{p} \}$ and the set of $L$ physical links $E^{p}=\{ e^{1}_{p}, \dots, e^{L}_{p} \}$. Each node $v^{i}_{p}$ has a limited amount of computing resources, denoted as $\pmb{CPU}(v^{i}_{p})$ while the bandwidth of link $e^{i}_{p}$ is $\pmb{BW}(e^{i}_{p})$. 

% We assume that a virtual embedding request (VNR) is also in the form of an un-directed graph $G_{v}=(V_{v}, E_{v})$. Similar to the physical network, for each VNR, there are $n$ virtual nodes 
% % $V_{v}=\{ v^{1}_{v},\dots,v^{n}_{v} \}$ 
% and $l$ virtual links.
% % $E_{v}=\{ e^{1}_{v}, \dots, e^{V}_{v} \}$.
% Each virtual node $v^{i}_{v}$ requires $\pmb{CPU}(v^{i}_{v})$ units of resource for its operation, and each virtual link $e^{i}_{v}$ consumes $\pmb{BW}(e^{i}_{v})$ units of bandwidth to handle the communication of its endpoints.

\subsection{System Model}
The physical network is represented as an undirected graph \(G_p = (V_p, E_p)\). The set of physical nodes is denoted as \(V_p\), and the set of physical links is denoted as \(E_p\). Each physical node has limited computing resources denoted as \(\text{CPU}(v)\), where \(v \in V_p\), and the bandwidth of each physical link is denoted as \(\text{BW}(e)\), where \(e \in E_p\).

We assume that a virtual embedding request (VNR) is also in the form of an undirected graph \(G_v = (V_v, E_v)\). Similar to the physical network, there are virtual nodes denoted as \(V_v\) and virtual links denoted as \(E_v\) for each VNR. Each virtual node requires resources denoted as \(\text{CPU}(v')\), where \(v' \in V_v\), and each virtual link consumes bandwidth denoted as \(\text{BW}(e')\), where \(e' \in E_v\).

As a decentralized VNE algorithm, there is no centralized controller in the system. Every physical node can serve the VNR, and clients can send their VNRs to any one of the physical nodes. It ensures that a single node failure will not halt the system since clients can resend the request to other physical nodes if they receive no response from the primary node. 

\subsection{Performance Metrics}
The primary objective of this study is to minimize the blocking probability of virtual networks, which plays a critical role in optimizing resource utilization, ensuring high service availability, and enhancing customer satisfaction. The acceptance ratio can be mathematically represented as:

\fontsize{8}{8}\selectfont
\begin{equation}
    \text{Acceptance Ratio} = 
    \lim_{\vert\text{VNRs}\vert\rightarrow \infty} \frac{\sum_{t\in\text{VNRs}} e_{t}}{\vert\text{VNRs}\vert}
\end{equation}
\normalsize

In this context, $e_t$ represents a binary variable that indicates whether a single VNR is embedded within the physical network or not.
Furthermore, the \emph{Revenue} and \emph{Cost} of the algorithm's embedding can be computed as follows:

\fontsize{8}{8}\selectfont
\begin{gather}
    \text{Revenue} 
    =
    \sum_{t\in\text{VNRs}}
    e_{t} \Big\{
    \sum_{v' \in V_v}
    \text{CPU}(v')
    +
    \sum_{e' \in E_v}
    \text{BW}(e')\Big\}, \label{eq_rev} \\
    \begin{split}
        \text{Cost} =
        \sum_{t\in\text{VNRs}} &
        e_{t} \Big\{
        \sum_{v' \in V_v}
        \text{CPU}(v') \\
        & + 
        \sum_{i, j \in V_p}
        \sum_{p_{i,j} \in \mc{P}_{i,j}}
        \sum_{e' \in E_v}
        y_{i,j}
        \text{BW}(e')\vert p_{i,j}\vert\Big\},
    \end{split}
    \label{eq_cost} 
\end{gather}
\normalsize

Here, the binary decision variable $y_{i,j}$ is defined to denote if the virtual link $e'$ is mapped onto the physical path $p_{i, j}$ between the physical nodes $i$ and $j$ or not. The symbol $\vert p_{i,j}\vert$ represents the length of the path $p_{i,j}$.

% In {\ourAlg}, we introduce a specific metric with the primary aim of minimizing this metric as much as possible. The optimization equation representing this metric is depicted as follows:

% \fontsize{8}{8}\selectfont
% \begin{gather}
%     \max_{X,Y} \: \text{metric} = \max_{X,Y} \: X \cdot \text{Revenue} - Y \cdot \text{Cost}
% \end{gather}
% \normalsize

\subsection{Algorithm}
When a physical node receives a Virtual Network Request (VNR), it acts as the primary node for that request and starts the decentralized Virtual Network Embedding (VNE) algorithm. We select a random list of physical nodes to act as leaders. These leaders are tasked with performing local embedding. This process generates embedding candidates, each with its own cost and revenue.
Then, we run a modified version of the ring-based election algorithm (explained in detail in Algorithm \ref{alg}) among the leader nodes. This helps us to collectively agree on the optimal embedding candidate with the highest metric score. Finally, the physical resources are allocated to the client in accordance with the selected proposal.

\begin{algorithm}[t]
  \fontsize{8}{8}\selectfont
  \caption{\ourAlg\ -- \ourAlgFull}
  \label{alg}
  \DontPrintSemicolon\SetNoFillComment
  \SetKwFunction{procedureName}{\textbf{\ourAlgName}} 
  \SetKwProg{myalg}{procedure}{}{}
  \nonl \myalg{\procedureName{%
    $G_{p}$, $G_{v}$, $X$, $Y$, $L$%
  }}{
    $P \leftarrow$ a random circular linked list of $L$ selected leaders in the physical network\;

    \tcc{Embedding Initiation: }
    cost, revenue, sol $\gets$ embed($n, G_{p}, G_{v}$)  \\

    $metric$ $\gets$ X $\times$ revenue $-$ Y $\times$ cost \\
    $p_n$ creates an EMBEDDING message with $id = (n, metric, P, G_p, G_v)$\;
    $p_n$ passes the EMBEDDING message to next node\;

    \tcc{Message Handling: }
    \For{$p_i \in P$}{
        \If{$p_i$ receives EMBEDDING message}{
            cost, revenue, sol $\gets$ embed($i, G_{p}, G_{v}$)  \\
            $metric_{pi}$ $\gets$ X $\times$ revenue $-$ Y $\times$ cost \\
            \If{message metric $>$ $metric_{pi}$}{
                $p_i$ forwards the EMBEDDING message to $p_{i+1}$
            }
            \Else{
                $p_i$ changes the EMBEDDING message id to $(i, metric_{pi}, P, G_p, G_v)$ and forwards the message to $p_{i+1}$\;
            }
        }
        
        \If{$p_i$ receives EMBEDDING message with its own id}{
            $p_i$ sends EMBEDDED message with $id = (i, sol, P, G_p, G_v)$ to $p_{i+1}$\;
        }
        
        \If{$p_i$ receives EMBEDDED message with its own id}{
            \tcc{No more messages, terminated}
            allocate($G_{p}$, $G_{v}$, sol) \label{alg_line_allocate} \\
        }

        \If{$p_i$ receives EMBEDDED message}{
            sol $\gets$ EMBEDDED message \\
            $p_i$ forwards the EMBEDDED message to $p_{i+1}$\;
        }

    }

}{}
\end{algorithm}

\cat{Decentralized Embedding} 
As shown in algorithm \ref{alg}, the primary node launches {\ourAlgName} using a virtual network request, $G_v$, along with the physical network, $G_{p}$. There are three adjustable parameters configured to align with the network provider's requirements: $(X,Y)$, which are used to balance revenue and cost in the embedding performance metrics, and $L$, which determines the number of leaders to be chosen.

In the embedding initialization phase, a list of $L$ leaders (including the primary node itself) $P$ is chosen randomly in the physical network by the primary node. We chose leaders randomly for each request because it can improve the utilization of physical nodes compared to having a fixed set of leaders. The primary node then performs a low-cost local embedding algorithm to find out an embedding candidate that can fit the VNR and calculates the metric score using the cost/revenue of the candidate. It generates an \textit{EMBEDDING} message to start the ring-based algorithm by sending the message to the next node in the leaders list $P$. Note that the list $P$ is constructed as a circular linked list.

In the first phase of the ring-based algorithm, the EMBEDDING message must go through all leaders in the list $P$ at least once to find the best embedding candidate. When a node $n_i$ receives an EMBEDDING message with an id $(j, metric_j, P, G_p, G_v)$ where $i \neq j$, it performs the local embedding algorithm and calculates the metric score $metric_i$. If $metric_i > metric_j$ in the message id, node $n_i$ has a better embedding solution, so it changes the message id to $(i, metric_i, P, G_p, G_v)$ and sends it to the next node. Otherwise, it simply forwards the original message indicating that it agrees with the current solution.

When a node $n_i$ receives an EMBEDDING message with its own solution, it knows that all leaders have agreed on its proposal. It then sent an EMBEDDED message with the solution to the next leader, which will again go through every leader to notify the final decision. The algorithm terminates when $n_i$ receives the EMBEDDED message created by itself. $n_i$ will allocate the resources for the VNR and reply to the client.

\cat{Local Embedding} We adopted the local embedding algorithm from previous work GraphViNE~\cite{habibi2020graphvine}. This algorithm uses a breadth-first search algorithm to find a possible embedding solution.
% , as shown in algorithm \ref{alg_embed}.
When a leader $n_i$ calls the algorithm, it starts searching with $n_i$ itself as the root node. Firstly, the virtual nodes $V_{v}$ are sorted as higher resource demand first queue. Then a breadth-first search algorithm is performed on the physical network $G_{P}$. Each traversed node will greedily embed as many virtual nodes as possible while meeting the CPU and bandwidth requirements. When all virtual nodes have been embedded in physical nodes, the embedding solution is found, and we compute the cost and revenue. To reduce the search space, two thresholds, $\alpha\times|V_v|$ and $\beta$ are defined to limit the total number of inspected servers and the maximum search depth, respectively. 

\section{Evaluation}
% TODO: fraction of physical nodes - physical node utilization - PDA?

We carry out simulations to showcase the efficacy of our suggested method in relation to VN acceptance ratio, revenue, cost, and utilization compared with other centralized algorithms. We employ the random graph model~\cite{randomGraph} for the formation of physical and virtual networks. 
% All the algorithms are executed in Python $3.6.9$. The computations are performed on a computer furnished with an Intel$^{\tiny{\textregistered}}$ Core\texttrademark,$i5-9500T$ processor operating at $2.2 - 3.7$~GHz and equipped with $8$~GB of RAM.

\subsection{Parameters for Simulation}
In this subsection, we present and clarify the parameters employed during the evaluation.

\cat{Physical Network} For the physical network modeling, we opt for a typical real-world setup featuring $1.2$ terabytes of RAM and a hundred processing and graphics cores similar to those of, respectively, Dell\texttrademark\ PowerEdge\texttrademark\ R$910$, two Intel$^{\tiny{\textregistered}}$ Xeon$^{\tiny{\textregistered}}$ Scalable processors post Hyperthreading, and a single Nvidia$^{\tiny{\textregistered}}$ GeForce$^{\tiny{\textregistered}}$ GT $330$. We then apply a normal distribution to instill diversity and heterogeneity. As a result, we consider a network of $100$ servers, where there is a $40\%$ likelihood of a direct physical link between each pair of servers. Each physical link is distinguished by its bandwidth (in Mbps), which is randomly selected from the normal distributions $\mc{N}(100, 400)$. Every physical node is defined by its CPU power (number of cores), memory capacity (in GBytes), and GPU power (number of cores), with these values chosen from the normal distributions $\mc{N}(100, 400)$, $\mc{N}(1200, 300)$, and $\mc{N}(100, 400)$, respectively.

\cat{Virtual Networks} The count of virtual nodes in each VNR is randomly picked from the interval $[4, 10]$. The probability of a virtual link existing between two virtual nodes stands at $0.7$. The bandwidth demand of virtual links adheres to the normal distributions $\mc{N}(10, 4)$. The CPU, memory, and GPU demand of each virtual node are derived from the normal distributions $\mc{N}(10, 4)$, $\mc{N}(30, 9)$, and $\mc{N}(10, 4)$, respectively. Every VN possesses a lifetime that is also randomly selected from a normal distribution of $\mc{N}(100, 900)$. VNs arrive according to the arrival rate and persist in the physical network for the span of their lifetime. The VN arrival rate is set at $2$ per unit of time and the simulation runs for $2000$ time units.

\cat{Algorithms for Comparison}
In conjunction with \ourAlg, we have executed the following algorithms to make a comparative study.
\begin{itemize}[leftmargin=*]
    \item \textbf{FirstFit}: This is an algorithm that embeds virtual nodes into the first available physical node with adequate capacity.
    \item \textbf{BestFit}: This algorithm opts for the physical node boasting maximum CPU capacity and fills it with the demands of the virtual node.
    \item \textbf{GRC}~\cite{grc}: An algorithm based on node-ranking.
    \item \textbf{NeuroViNE}~\cite{neurovine}: This algorithm uses a search space reduction mechanism. It extracts pertinent subgraphs by means of a Hopefield network, and then employs GRC to embed VN into candidate subgraphs.
\end{itemize}
The parameters $\alpha$, $\beta$, and $L$ are set to $30$, $3$, and $5$ respectively. The optimization parameters $X$ and $Y$ are both set to $1$.

\subsection{Benchmarks}
In this subsection, we benchmark our algorithm against centralized algorithms in terms of performance to demonstrate that {\ourAlg} can be utilized in real-world applications with acceptable results. It's important to note that this algorithm is decentralized, thus it benefits from all the advantages detailed in section~\ref{sec:motivation}.

\begin{figure}[t]
	\centering
	\includegraphics[width=0.97\linewidth]{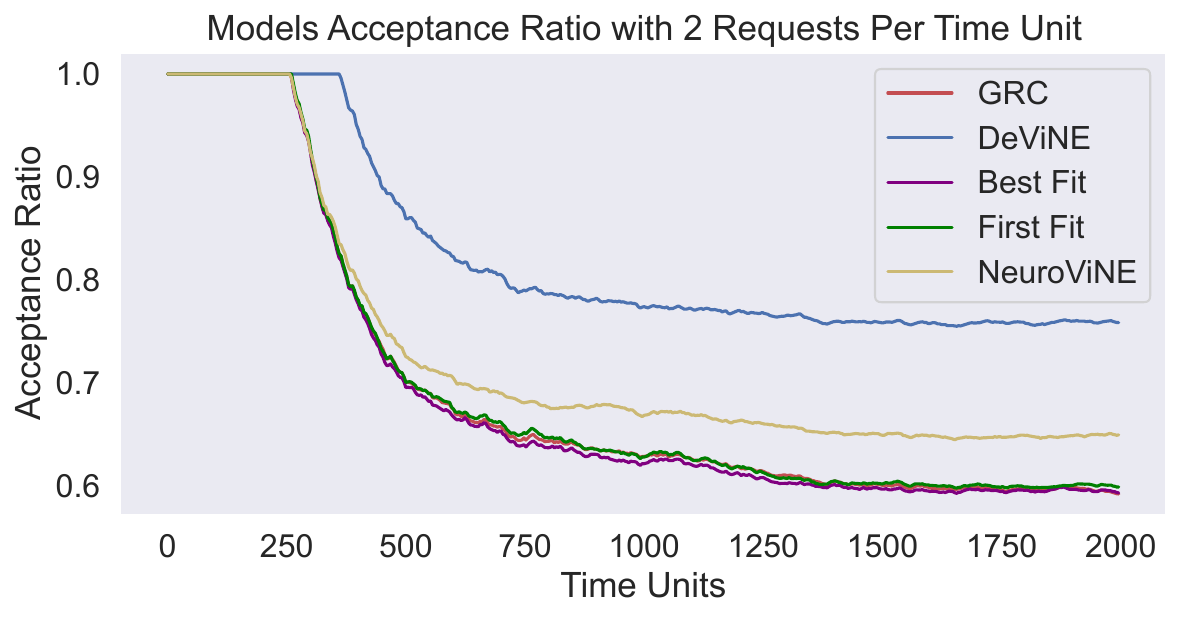}
	\caption{Acceptance Ratio over Time}
	\label{fig:ar}
\end{figure}
\cat{Acceptance Ratio}
The long-term acceptance ratio stands as a significant metric influencing the system's profitability. As depicted in Figure~\ref{fig:ar}, the acceptance ratio of varying algorithms over a lengthy simulation of approximately $2000$ episodes is displayed. It should be noted that during this period, the acceptance ratios attain a state of equilibrium and maintain consistency due to a stationary virtual network arrival process. \ourAlg\ enhances the acceptance ratio by roughly $12\%$ in comparison to NeuroViNE and by $17\%$ in relation to other algorithms.

\begin{figure}[t]
    \begin{subfigure}{0.49\linewidth}
	\centering
	\includegraphics[width=\linewidth]{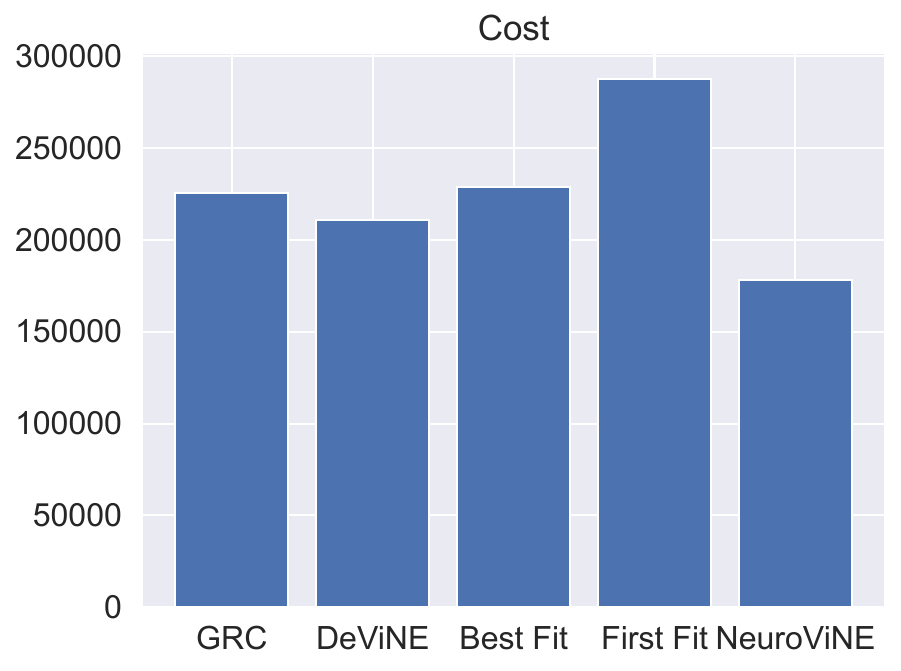}
	\caption{Cost of Embedding}
	\label{fig:cost}
    \end{subfigure}
    \hfill
    \begin{subfigure}{0.49\linewidth}
	\centering
	\includegraphics[width=\linewidth]{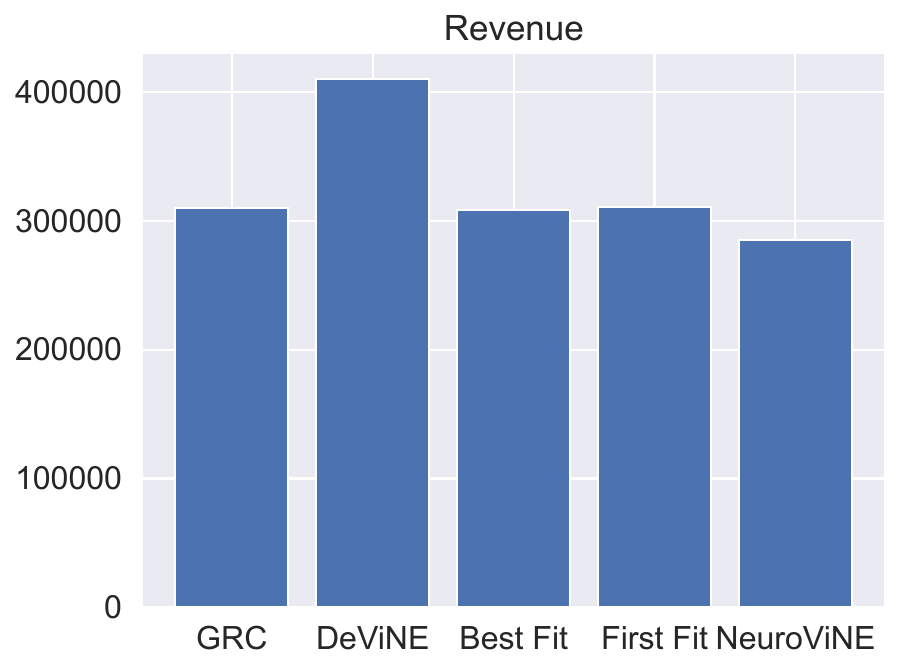}
	\caption{Revenue of Embedding}
	\label{fig:revenue}
    \end{subfigure}
    \caption{Financial Performance of \ourAlg}
    \label{fig:fperformance}
\end{figure}
\cat{Revenue and Cost}
Figures~\ref{fig:cost} and \ref{fig:revenue}, respectively, compare the cost and revenue of different algorithms. Computations are based on the equations~\eqref{eq_rev} and \eqref{eq_cost}.
Given that \ourAlg\ accommodates more VNs over the same duration, it generates a higher revenue. Even with a higher revenue, our proposed algorithm results in a lower cost compared to the First Fit, Best Fit, and GRC methods. NeuroViNE manages a low cost because it typically embeds smaller virtual networks, and as a result, it cannot generate high revenue. Furthermore, \ourAlg\ achieves the maximum revenue-to-cost ratio (\ie\ $1.94$), while NeuroViNE, GRC, Best Fit, and First Fit achieve ratios of $1.6$, $1.37$, $1.34$, and $1.08$, respectively.

\begin{figure}[t]
    \begin{subfigure}{0.49\linewidth}
	\centering
	\includegraphics[width=\linewidth]{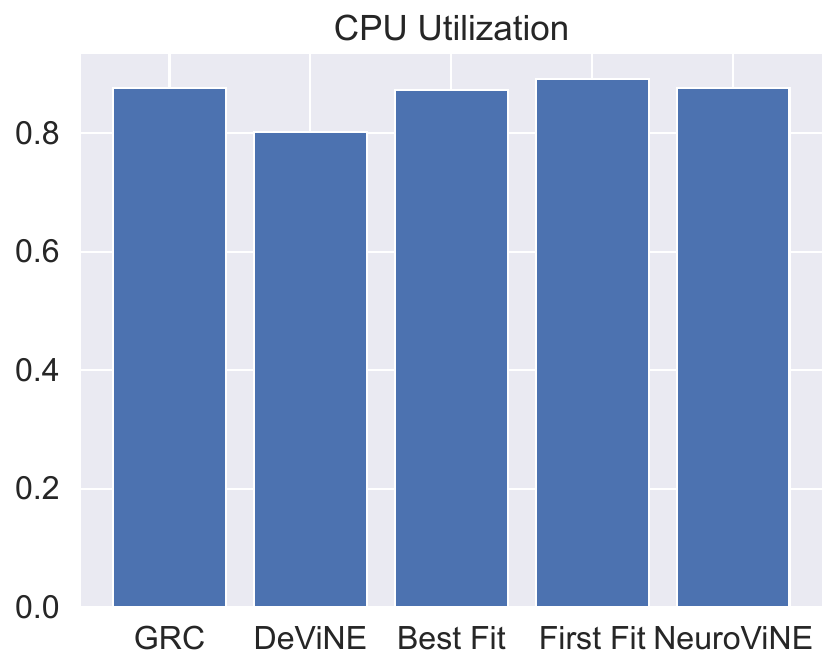}
	\caption{Average CPU Utilization}
	\label{fig:cpu-util}
    \end{subfigure}
    \hfill
    \begin{subfigure}{0.49\linewidth}
        \centering
	\includegraphics[width=\linewidth]{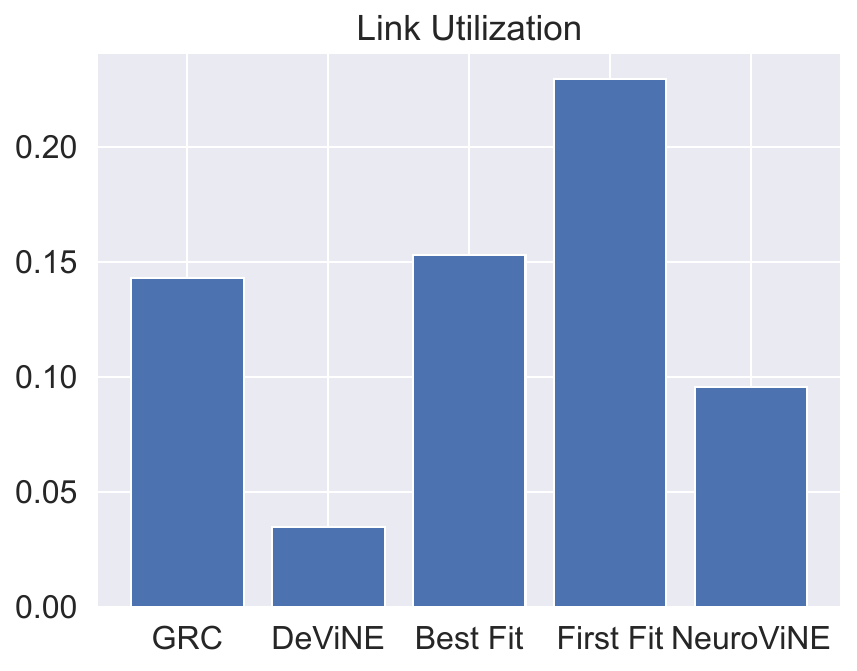}
	\caption{Average Link Utilization}
	\label{fig:link-util}
    \end{subfigure}
    \caption{Utilization of \ourAlg}
    \label{fig:utilization}
\end{figure}
\cat{Utilization}
Figure~\ref{fig:cpu-util} and \ref{fig:link-util} represent CPU and link utilization in comparison to other methodologies. \ourAlg~ outperforms others in terms of link utilization. This is due to \ourAlg~ employing a BFS-like algorithm for embedding and restricting the depth of BFS, which results in the usage of fewer links. \ourAlg~ also shows slightly lower CPU utilization than other methods as the leaders are chosen randomly. However, CPU utilization can be enhanced by augmenting the number of leaders in each embedding round.

\subsection{Complexity}
% In this subsection, we discuss the complexity of our algorithm in terms of communication and computation. 

\cat{Computation Complexity}
The computational complexity of {\ourAlg} is determined by the local embedding algorithm, which is executed for each leader in the embedding process. The local embedding algorithm has a time complexity equivalent to the BFS algorithm, with a maximum number of inspected nodes denoted as $\alpha\times|V_v|$. Consequently, the overall computational complexity of {\ourAlg} can be expressed as $O(L\times\alpha\times|V_v|)$, which demonstrates its efficiency compared to many centralized methods.

\cat{Communication Complexity}
The decentralized nature of {\ourAlg} introduces a communication complexity to the algorithm. In the worst-case scenario, the communication complexity is determined by the number of messages exchanged between the servers. Specifically, it involves sending $2L - 1$ \textit{EMBEDDING} messages when the second-to-last server possesses the best local embedding and $L$ \textit{EMBEDDED} messages for notifying all the servers. Thus, the communication complexity of {\ourAlg} can be shown as $O(3L - 1)$.
\section{Conclusion}

In this study, we introduced \ourAlg~, a decentralized Virtual Network Embedding (VNE) algorithm. The decentralized approach to VNE offers superior scalability, robustness, and security in comparison to prior algorithms.
We benchmarked our algorithm against existing heuristic and preprocessing algorithms, demonstrating that our method maintains strong performance despite its fully decentralized nature.
%
%TODO: future work

% \IEEEtriggeratref{16}
\bibliographystyle{IEEEtran}
\bibliography{cite}

\end{document}